\documentclass[sn-mathphys-num]{sn-jnl}

\usepackage{graphicx}%
\usepackage{multirow}%
\usepackage{amsmath,amssymb,amsfonts}%
\usepackage{amsthm}%
\usepackage{mathrsfs}%
\usepackage[title]{appendix}%
\usepackage{xcolor}%
\usepackage{textcomp}%
\usepackage{manyfoot}%
\usepackage{booktabs}%
\usepackage{algorithm}%
\usepackage{algorithmicx}%
\usepackage{algpseudocode}%
\usepackage{listings}%
\usepackage{array}
\usepackage{subfigure}
\usepackage{fancybox}
\usepackage{pifont}
\usepackage{tablefootnote}
\usepackage[export]{adjustbox}

\usepackage{geometry}
\usepackage{rotating}
\usepackage{soul}
\usepackage{amssymb}
\usepackage{pifont}
\usepackage{dsfont}

\definecolor{darkblue}{rgb}{0.0, 0.0, 0.55}
\theoremstyle{thmstyleone}%

\theoremstyle{thmstyletwo}%

\theoremstyle{thmstylethree}%

\begin{document}

\title{Disentangling multispecific antibody function with graph neural networks}

\author[1]{Joshua Southern}
\author[1]{Changpeng Lu}
\author[1]{Santrupti Nerli}
\author[1]{Samuel D. Stanton}
\author[1]{Andrew M. Watkins}
\author[1]{Franziska Seeger}
\author[1]{Fr\'ed\'eric A. Dreyer}

\affil[1]{\orgdiv{Prescient Design}, \orgname{Genentech}, \orgaddress{\city{South San Francisco}, \postcode{CA}, \country{USA}}}

\abstract{
Multispecific antibodies offer transformative therapeutic potential by engaging multiple epitopes simultaneously, yet their efficacy is an emergent property governed by complex molecular architectures. Rational design is often bottlenecked by the inability to predict how subtle changes in domain topology influence functional outcomes, a challenge exacerbated by the scarcity of comprehensive experimental data. Here, we introduce a computational framework to address part of this gap. First, we present a generative method for creating large-scale, realistic synthetic functional landscapes that capture non-linear interactions where biological activity depends on domain connectivity. Second, we propose a graph neural network architecture that explicitly encodes these topological constraints, distinguishing between format configurations that appear identical to sequence-only models. We demonstrate that this model, trained on synthetic landscapes, recapitulates complex functional properties and, via transfer learning, has the potential to achieve high predictive accuracy on limited biological datasets. We showcase the model's utility by optimizing trade-offs between efficacy and toxicity in trispecific T-cell engagers and retrieving optimal common light chains. This work provides a robust benchmarking environment for disentangling the combinatorial complexity of multispecifics, accelerating the design of next-generation therapeutics.
}


\maketitle

\section{Introduction}\label{sec:intro}

Advances in antibody engineering have paved the way for a therapeutic revolution, moving beyond traditional monoclonal antibodies  to more complex and potent molecular formats~\cite{runcie2018bi, yao2023trispecific}. Bispecific and multispecific (msAb) antibodies, capable of simultaneously binding two or more distinct epitopes, are at the forefront of this new generation of biologics. This capability unlocks novel mechanisms of action previously unattainable, such as redirecting T-cells to tumors, concurrently blocking redundant signaling pathways, or enhancing specificity toward target cells~\cite{chames2009bispecific, tapia2023bi}. The clinical success of early msAbs, such as the bispecific T-cell engager (TCE) blinatumomab and the trifunctional antibody catumaxomab, demonstrated remarkable efficacy and established msAbs as a transformative class of therapeutics~\cite{chames2009bispecific, krishnamurthy2018bispecific}.

The field has rapidly evolved to produce a diverse arsenal of msAb formats, including trispecific and even tetraspecific constructs~\cite{castoldi2016tetramabs}, engineered from building blocks like single-chain variable fragments (scFv) and Fabs~\cite{runcie2018bi, wu2019building, HUDSON1999177}. Innovation has even extended to the domain level, with the development of dual-targeting Fabs that engineer two spatially separated paratopes within a single Fv region to engage two targets simultaneously~\cite{beckmann2021dutafabs}. 
These sophisticated molecules enable multifaceted therapeutic strategies. Dominant approaches include immune cell engagers that bridge T-cells or Natural Killer cells to tumors~\cite{ochoa2017antibody, gleason2012bispecific}, dual-targeting of tumor-associated antigens to prevent immune escape~\cite{zhao2022novel, kugler2010recombinant}, and providing both primary (e.g., via CD3) and co-stimulatory (e.g., via CD28) signals to achieve robust T-cell activation~\cite{wu2020trispecific, seung2022trispecific, mullard2020trispecific}.
Emerging strategies also include leveraging myeloid cells within the tumor microenvironment; for instance, by engaging PD-L1 on macrophages to induce a feed-forward loop of T-cell activation~\cite{yang2025trispecific}.
Other strategies involve the simultaneous blockade of multiple immune checkpoint pathways to overcome tumor-induced immunosuppression~\cite{yao2023trispecific, kalbasi2020tumour, xu2020regulation}. While early formats were often assembled via chemical cross-linkers like SPDP~\cite{yao2023trispecific, somasundaram1999development}, modern recombinant technologies have enabled the creation of numerous stable and clinically viable architectures~\cite{tapia2023three}.

Despite this promise, the rational design of msAbs presents a profound challenge. The therapeutic function of an msAb is not merely the sum of its parts; it is an emergent property of its specific molecular architecture. Variations in binding valency and the geometric arrangement of binding domains can dramatically alter functional outcomes like potency, specificity, and safety~\cite{zhao2022novel, dicara2022development}. This creates a vast combinatorial design space, reflected in the over 100 distinct molecular formats currently navigating clinical development~\cite{brinkmann2026making}. 
In this expanding landscape of architectures, subtle changes in the positioning of a binding domain can lead to unexpected and non-additive effects on biological activity. For instance, a trispecific construct may exhibit synergistic potency~\cite{dimasi2015development}, while another might suffer from steric hindrance or trigger unintended T-cell activation~\cite{jung1991target}.

Recent experimental studies have further illuminated this structure-function dependency, particularly for TCEs, which now constitute nearly 40\% of all multispecifics in clinical trials~\cite{brinkmann2026making}. For example, in the design of trispecific TCEs, moving a high-affinity binding domain from a distal to a proximal position relative to the Fc core was shown to completely decouple anti-tumor efficacy from fatal cytokine release syndrome in \textit{in vivo} models~\cite{dicara2022development}. Similarly, the precise geometry and flexibility of the linker sequences have been identified as critical determinants of immunological synapse formation; a recent study combining small-angle X-ray scattering with functional assays demonstrated that rigidifying the synapse can enhance potency even when affinity is constant~\cite{leithner2025solution}. These topological effects extend beyond simple potency. Variations in valency, such as the `2+1' format utilized in CEA-targeting bispecifics, have been shown to drastically alter biodistribution and synapse stability compared to `1+1' architectures~\cite{bacac2016cea, boje2024impact}. Furthermore, optimization of the cytokine window through geometric tuning, rather than just affinity attenuation, has emerged as a viable strategy to widen the therapeutic index~\cite{zhou2024using}.

Evaluating this enormous design space experimentally is a primary bottleneck. Functional assays such as \textit{in vitro} cytotoxicity, often measured by EC50, must be performed on the final, complex-format molecules. These assays are resource-intensive, noisy, and prohibitively expensive to generate at scale. Furthermore, the development of msAbs is complicated by manufacturing complexities and significant safety concerns, most notably cytokine release syndrome and on-target, off-tumor toxicity \cite{cao2025clinical, yao2023trispecific}. To circumvent these hurdles, researchers often measure simpler properties, like binding affinity kinetics from surface plasmon resonance, on the parental mAb formats. However, this parental binding data frequently fails to correlate with the functional activity of the final, engineered msAb, whose properties are highly format-dependent. This results in a severe scarcity of the rich, multi-format functional data truly needed to train robust predictive models.

Current computational approaches struggle to navigate this landscape efficiently. While physics-based methods such as molecular dynamics and coarse-grained simulations have been successfully employed to model specific developability attributes like viscosity~\cite{lai2022machine, anapindi2025leveraging} and aggregation~\cite{siegmund2025optimizing}, or to simulate immune synapse formation~\cite{su2024computational, ray2024mechanistic}, they are often too computationally expensive for high-throughput screening of the vast combinatorial design space. Conversely, traditional sequence-based machine learning models typically treat the antibody as a linear string, failing to capture the complex, non-local 3D interactions that define multispecific function. Recently, deep learning has begun to bridge this gap. Generative protein language models and active learning frameworks have achieved remarkable efficiency in affinity maturation and developability optimization for monoclonal antibodies and single domains~\cite{hie2024efficient, li2023machine, frey2025lab, gruver2023protein, wang2025guided, sinha2025cdr}. Furthermore, the integration of structural priors has proven capable of traversing complex binder landscapes and rescuing binding against viral escape variants~\cite{shanker2024unsupervised, dreyer2025computational}. Yet, with the notable exception of the EVA platform~\cite{grace2025engineering}, the majority of recent deep learning advances, including developability profiling~\cite{raybould2019five} and generative foundation models~\cite{wang2025iggm}, remain focused on the optimization of variable regions within standard monoclonal formats. Consequently, models that rely solely on the sequence of constituent variable domains typically fail to distinguish between different molecular formats. 
This limitation is increasingly critical as the field shifts toward asymmetric architectures, such as the emerging `2+1' formats used in nearly 15\% of clinical TCEs, where efficacy is governed by the specific geometric arrangement of valencies~\cite{brinkmann2026making}.
Furthermore, computational strategies that simply treat the parental binding affinities of individual domains as an additive proxy for the complex functional EC50 are imperfect. This assumption of additivity breaks down as it cannot capture the non-additive, structure-dependent properties, such as avidity gating and steric shielding, that govern msAb function~\cite{zhou2025ai}. Therefore, an important goal in computational antibody engineering is to develop models that can predict complex functional outcomes by integrating information from both the constituent binding domains and their specific molecular connectivity.

To overcome the shortage of comprehensive functional data for multispecific architectures, we introduce a dual computational framework. We first introduce a generative method using a novel extension of Ehrlich Functions~\cite{chen2025generalists} to create large, tunable datasets that mimic the complex, asymmetric, and non-linear nature of msAb functional data. We then use this data to train a Graph Neural Network (GNN) that explicitly represents antibody formats as graphs, enabling the prediction of holistic functional properties from domain-specific inputs. We demonstrate that this model can be trained on our synthetic data to recapitulate complex functional properties, as well as achieve high predictive accuracy on small biological datasets through transfer learning of a pretrained model. This work provides a powerful, topology-aware computational framework for accelerating the rational design of next-generation multispecific therapeutics.

\section{Synthetic Data and Modeling Framework}
\label{sec:framework}

To study the relationship between molecular geometry and therapeutic function, we establish a benchmarking environment composed of two integrated modules. We first describe the synthetic data generation pipeline, Synapse (Synthetic Network of Antibody Protein Structure Elements), which simulates non-linear, epistatic interactions. We then define the accompanying topology-aware learning architecture used to disambiguate these combinatorial complexities

\begin{figure}
    \centering
    \includegraphics[width=1.0\linewidth]{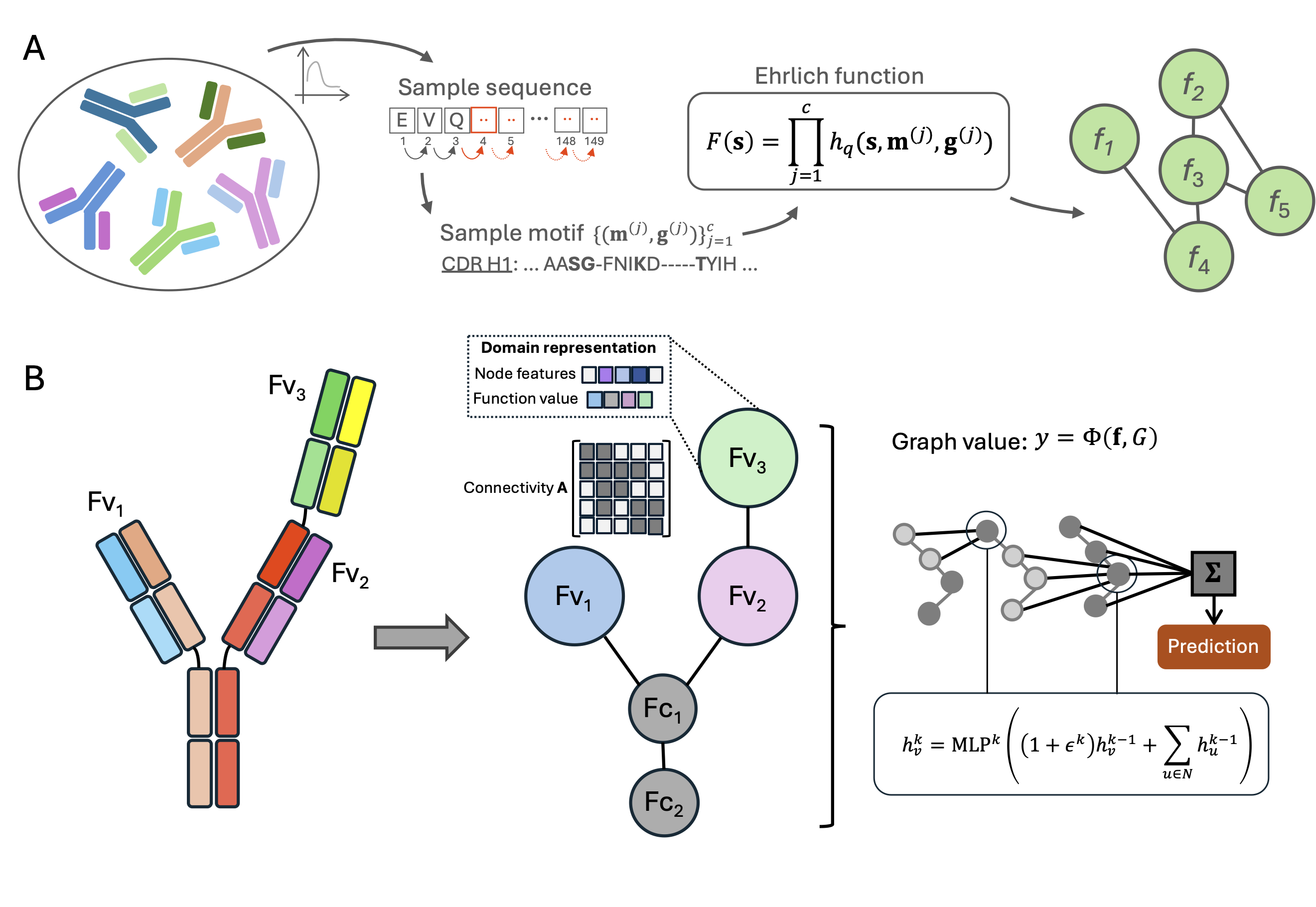}
    \caption{Overview of the Synapse data generation framework. \textbf{(A)} Realistic domain sequences are sampled from an OAS-derived PSSM. Intrinsic fitness scores are assigned via Ehrlich functions to mimic non-linear binding landscapes.
    \textbf{(B)} A multispecific antibody is represented as a graph. The final regression objective is derived from a global readout function that integrates intrinsic domain values with their specific topological connectivity, creating a ground-truth value where structure dictates function.}
    \label{fig:overview}
\end{figure}

\subsection{Synapse: A Graph-Based Synthetic Landscape}
\label{sec:synapse}
To bridge the data gap, we develop Synapse, a generative framework that simulates ground-truth functional data for arbitrary antibody formats, as shown in Figure~\ref{fig:overview}A. Synapse is built on the premise that the global function of an msAb is an emergent property of its graph structure, where nodes are binding domains and edges represent their physical connectivity. We assume that the specific sequences of constant regions are less predictive than the geometric topology they enforce, and therefore model constant domains through simplified zero-valued nodes.

The framework utilizes a novel graph-based extension of Ehrlich functions~\citep{chen2025generalists}. While originally designed for single-sequence optimization, we adapt these functions to capture the complex non-linear and epistatic nature of multi-domain biophysical fitness landscapes, assigning an intrinsic Ehrlich function to each binding domain. We further enforce biophysical plausibility by constraining domain generation using a position-aware statistical model derived from the Observed Antibody Space (OAS)~\citep{oas1,oas2}, ensuring synthetic sequences mirror empirical antibody amino acid distributions.

The global biological activity is then defined via a connectivity-dependent readout function, as detailed in Section~\ref{sec:methods-data}. Rather than summing isolated domain scores, this function explicitly models how a domain's contribution can be chemically and sterically influenced by its neighbors. This formulation allows us to simulate physical phenomena such as steric shielding or avidity, linking antibody sequences and graph architectures to a complex, non-linear ground truth, providing a robust testbed for model benchmarking.

\subsection{Graph Neural Networks for Function Prediction}
\label{sec:graph-model}
We formulate msAb property prediction as a graph regression task mapping an antibody graph $\mathcal{G} = (\mathcal{V}, \mathcal{E})$ to a scalar functional value $y$. We hypothesize that sequence composition alone is insufficient for prediction and that explicit encoding of the adjacency matrix $A$ is required to resolve format-dependent activity.

GNNs have achieved great success in learning tasks with pairwise dependencies~\citep{Scarselli, gori, micheli}. Typically, GNNs are based on the message-passing paradigm~\citep{gilmer2017neural} in which node features are aggregated over their local neighborhood recursively. This mechanism can allow the model to encode topological context, such as the distance of a variable domain from the Fc core or its relative positioning (e.g., distal vs. proximal) within a chain, which sequence-only models cannot resolve. As detailed in Section~\ref{sec:methods-gnn}, we employ a Graph Isomorphism Network (GIN)~\citep{xu2018how} in our experiments, selected as a prototypical message-passing neural network that matches the expressive power of the $1$-WL test.

To isolate the specific contribution of molecular topology to predictive accuracy, we benchmark the GIN against a connectivity-agnostic Multi-Layer Perceptron (MLP) baseline described in Section~\ref{sec:methods-mlp}. The MLP receives the identical set of domain features but is connectivity invariant, treating the molecule as an unordered set of domains. Consequently, performance divergence between the GIN and MLP serves as a direct metric for the extent to which biological function is governed by domain connectivity versus simple composition.

\section{Results}
\label{sec:results}

\subsection{Scaling with data size and format complexity}
\label{sec:scaling}

\begin{figure}
    \centering
    \begin{minipage}{0.74\textwidth}
    \includegraphics[width=1.0\linewidth]{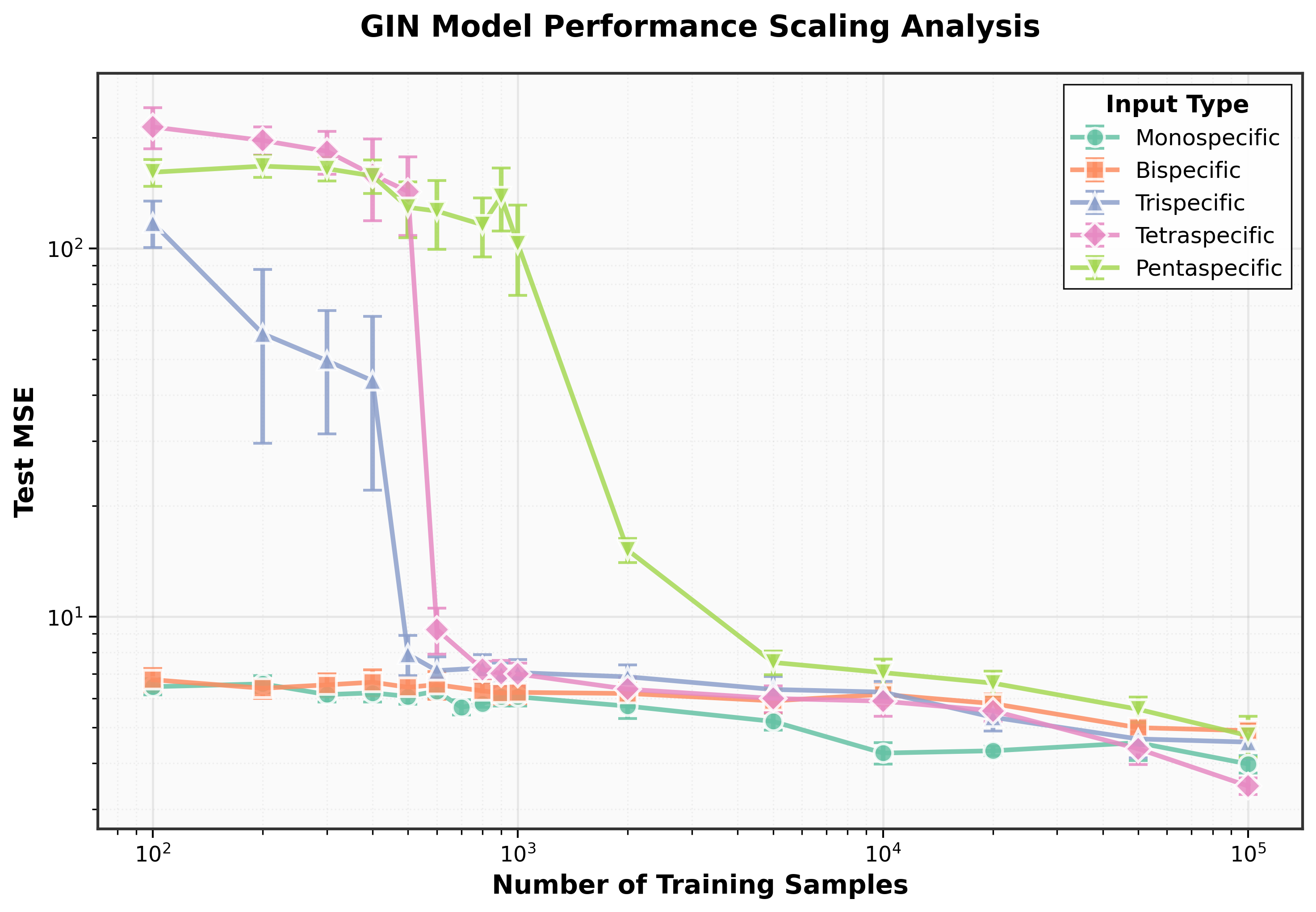}
    \includegraphics[width=1.0\linewidth]{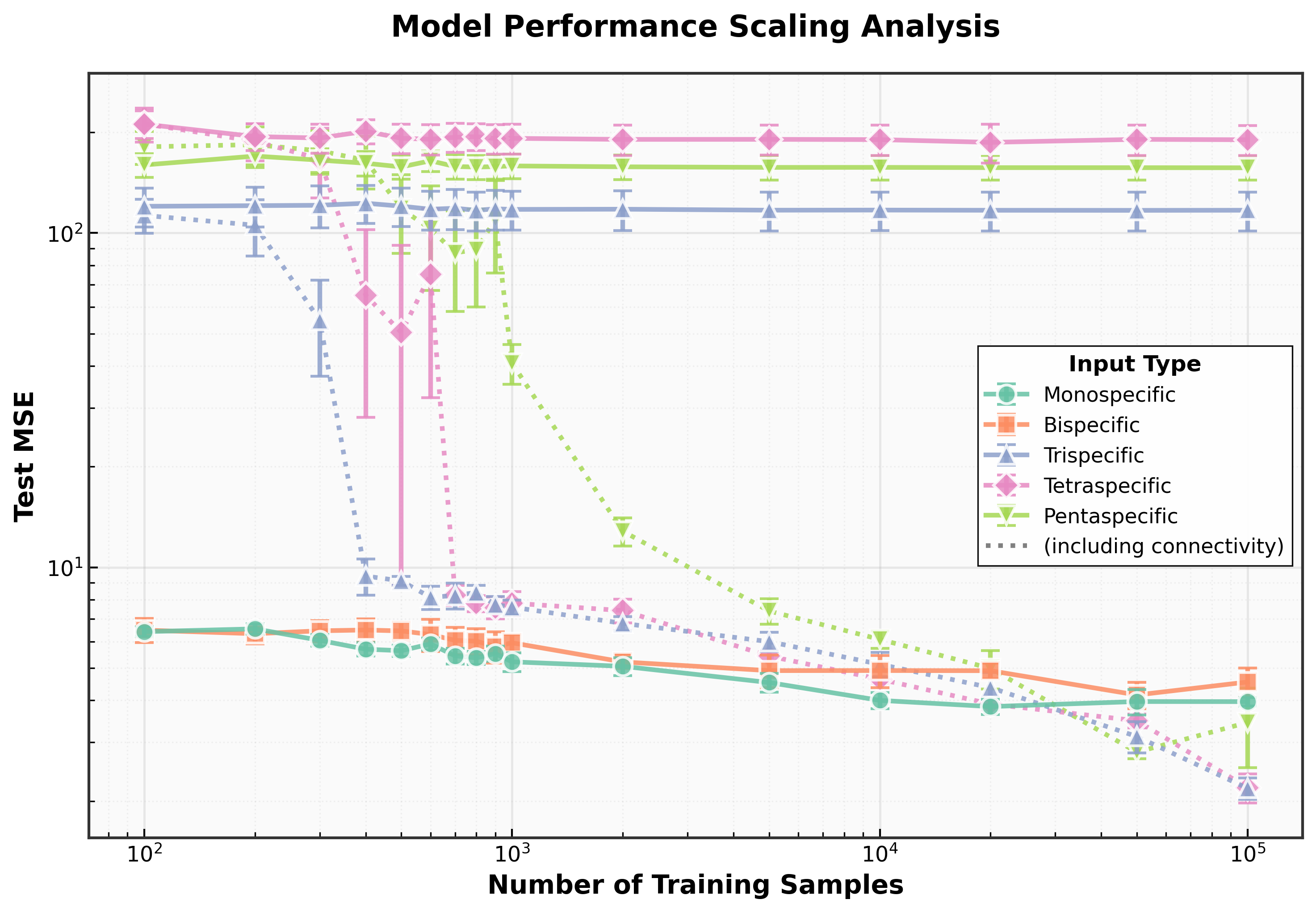}%
    \end{minipage}\hfill%
    \begin{minipage}{0.24\textwidth}
    \includegraphics[width=1.0\linewidth]{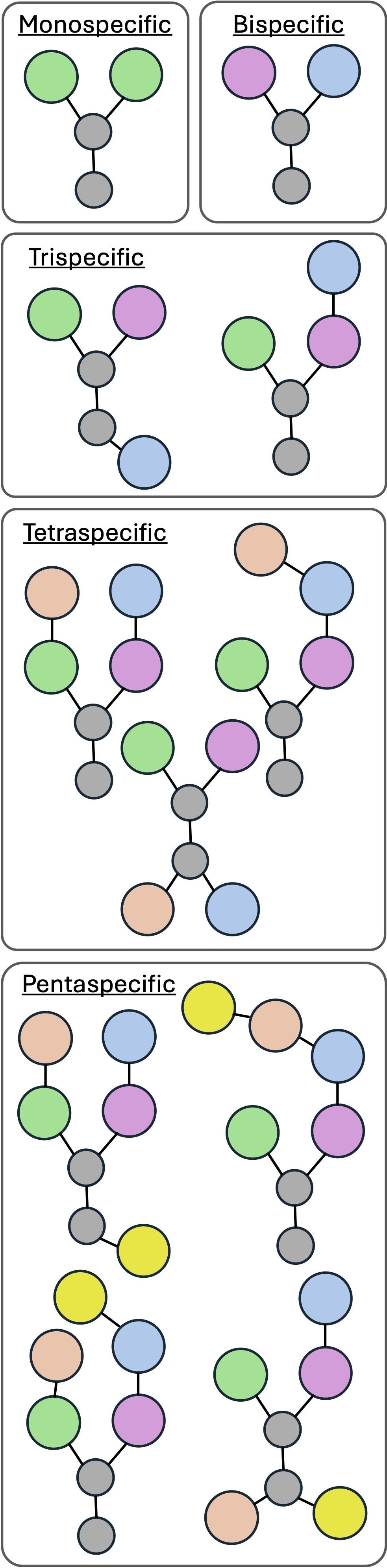}%
    \end{minipage}
    \caption{Scaling of model performance as a function of dataset size, for increasing format complexity, comparing a GIN to a MLP. Selected examples of possible domain connectivities are shown for each complex format.}
    \label{fig:scaling}
\end{figure}

To evaluate the impact of explicitly modeling molecular connectivity, we generated a comprehensive synthetic dataset using the Synapse framework. The dataset covers five levels of complexity, ranging from monospecific to pentaspecific antibodies. The graph generation process mimics biological constraints of IgG-like constructs: all formats share central Fc nodes with two primary variable arms. Additional variable domains connect to these arms or available constant nodes, creating a diverse set of branched structures. While Figure~\ref{fig:scaling} displays representative examples, the full dataset samples the complete space of valid connectivities for each format type.

Ground-truth functional values were calculated using Equation~(\ref{eq:scaling_phi}). This function sums weighted contributions from immediate (1-hop) and second-order (2-hop) neighbors. Consequently, the resulting global value depends on the specific spatial arrangement of the domains, rather than simply the list of domains present in the molecule.

We compared the GIN against the connectivity-unaware MLP baseline. Both models were trained on dataset sizes ranging from $10^2$ to $10^5$ samples and evaluated on a held-out test set, as shown in Figure~\ref{fig:scaling}.

For monospecific and bispecific formats, the GIN and MLP exhibited similar performance. In these simpler architectures, the single possible configuration allows the MLP to approximate the function effectively. However, a performance gap appeared for trispecific, tetraspecific, and pentaspecific formats.

As complexity increases beyond two variable domains, the MLP performance plateaus. The MLP processes the input as an unordered set of domains; therefore, it cannot distinguish between formats that contain the same variable domains arranged in different positions. Since the ground-truth function is connectivity-dependent, the MLP is unable to reduce error beyond a certain threshold. In contrast, the GIN explicitly encodes the adjacency matrix, allowing it to resolve these structural differences and achieve consistent performance improvements as dataset size increases.
As an additional baseline, we included an MLP model to which connectivity information is provided as a one-hot encoded vector of possible formats, showing that explicitly identifying the topology allows the MLP to perform comparably to the GNN. However, this approach is restricted to the fixed set of architectures defined in the training set, whereas the GNN can learn generalized topological rules that could theoretically extend to novel, unseen formats.

\subsection{Topology-Dependent Optimization of Trispecifics}
\label{sec:trispecific}

To demonstrate the capacity of our graph-based framework to capture non-linear, format-dependent biological phenomena, we instantiated a case study inspired by the development of Dual-Antigen Targeted T-cell Engagers (DAT-TCEs)~\cite{dicara2022development}. These trispecific molecules are designed to simultaneously engage T-cells via CD3 and tumor cells via two distinct antigens, Ly6E and B7-H4, to drive potent tumor eradication. In this biological system, Ly6E and B7-H4 were identified as ideal pairs for avidity-driven targeting due to their frequent co-expression on breast tumors while maintaining restricted, orthogonal expression profiles in healthy tissues. 

However, the development of such therapeutics presents a complex multi-objective optimization challenge where the therapeutic index is also governed by molecular geometry and not just affinity. Experimental data revealed that while high-affinity binding to B7-H4 is necessary for potency, it drives severe on-target, off-tumor toxicity in B7-H4-expressing tissues, such as the liver and uterus, when employed in traditional formats. In vivo studies demonstrated that this toxicity is topology-dependent: trispecific constructs presenting the high-affinity B7-H4 domain in a distal orientation relative to the Fc core precipitated rapid weight loss and pathology, whereas proximal positioning of the same domain retained anti-tumor efficacy while mitigating systemic adverse effects.

We simulated this landscape by defining a ground-truth global function $\Phi$ that formalizes these biological constraints. The functional score $y$ was calculated as a weighted sum of synergy, co-expression, and toxicity terms, conditioned explicitly on the graph structure $\mathcal{G}$
\begin{equation}
\label{eq:trispecific_function}
y = (w_{\text{syn},A} \delta_{\mathcal{G}_A} + w_{\text{syn},B} \delta_{\mathcal{G}_B})f_1 f_2 f_3 
+ w_{\text{co}}f_1 f_3 - w_{\text{tox},A} \delta_{\mathcal{G}_A} f_{\text{3}} - w_{\text{tox},B} \delta_{\mathcal{G}_B} f_{\text{1}}\,,
\end{equation}
where $\delta_{\mathcal{G}}$ is an indicator function that is 1 when the input graph is $\mathcal{G}$ (0 otherwise), and $f_i$ represents the intrinsic Ehrlich score of the constituent domains.
The coefficients are assigned to mimic the empirical observations: a ``safe'' topology (proximal B7-H4) is assigned moderate synergistic potency ($w_{\text{syn}}=0.2$) with a minimal toxicity penalty ($w_{\text{tox}}=0.1$), reflecting a favorable therapeutic window. Conversely, a ``toxic'' topology (distal B7-H4) is assigned higher potential potency ($w_{\text{syn}}=0.225$) but incurs a severe toxicity penalty ($w_{\text{tox}}=6$), rendering it therapeutically inviable. The co-expression term for the presence of anti-B7-H4 and Ly6E domains is set to $w_{\text{co}}=0.2$ for both formats.

\begin{figure}
    \centering
    \begin{minipage}{0.11\textwidth}
    \includegraphics[width=1.0\linewidth]{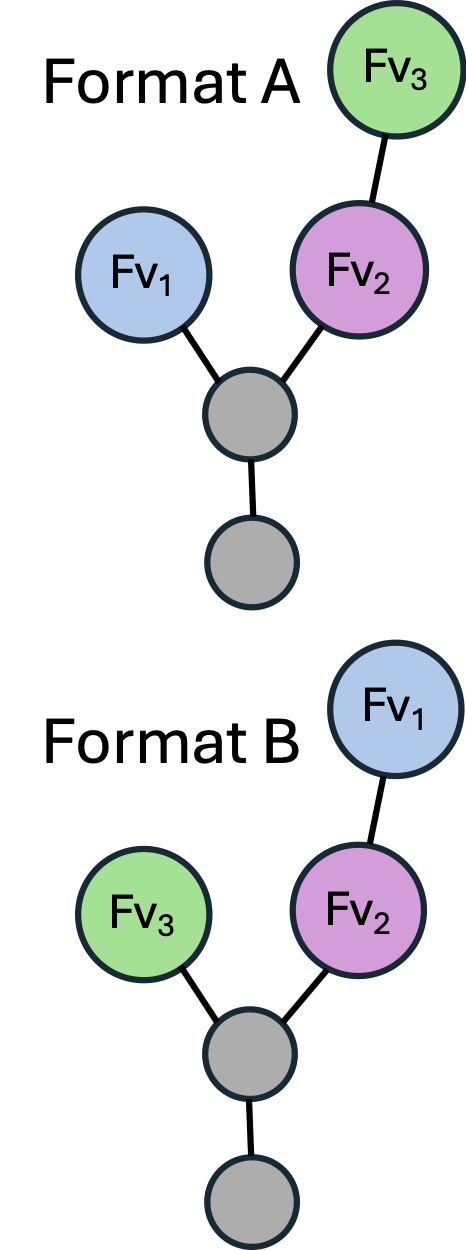}%
    \vspace{8mm}
    \end{minipage}\hfill%
    \begin{minipage}{0.435\textwidth}
    \includegraphics[width=1.0\linewidth]{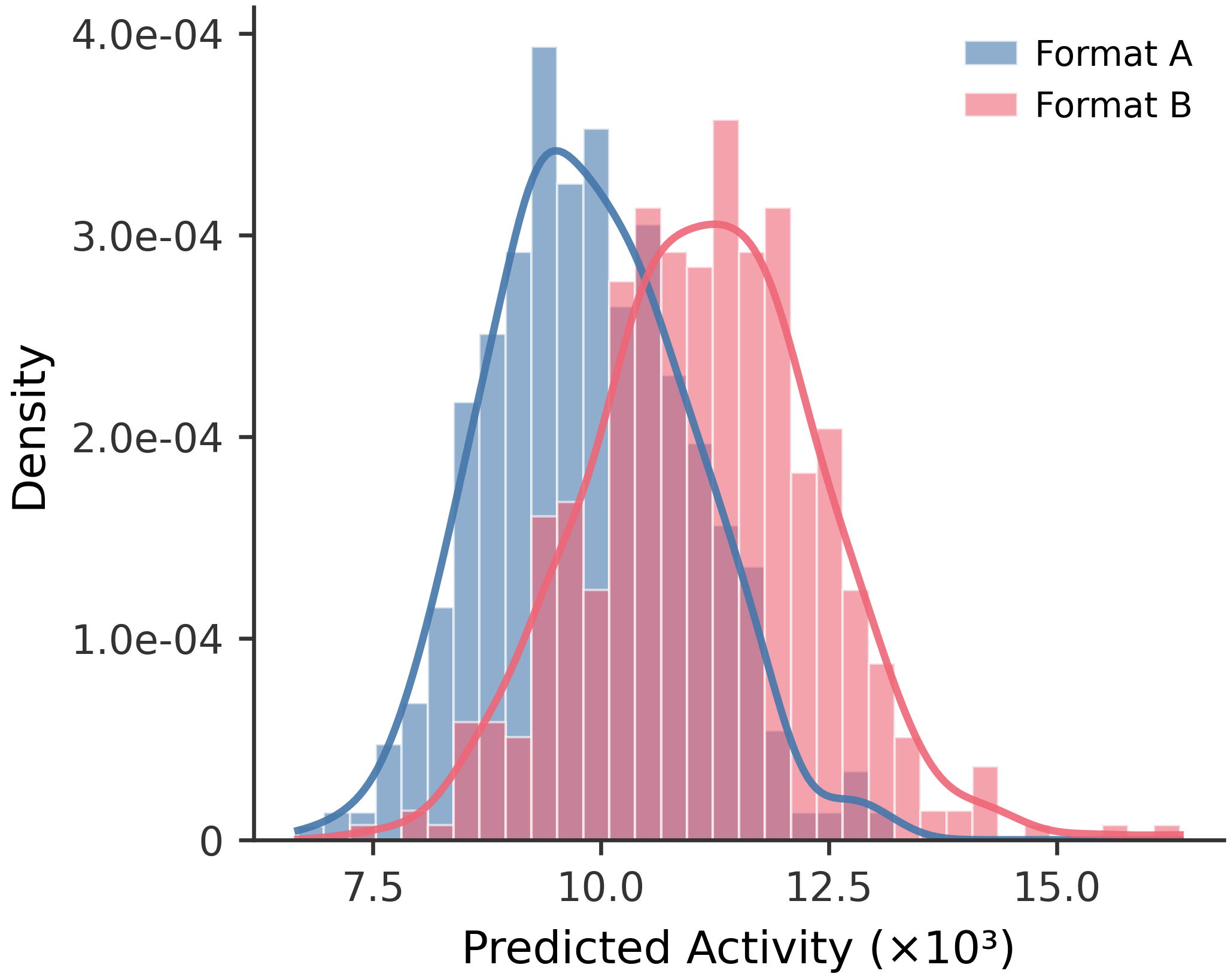}%
    \end{minipage}%
    \begin{minipage}{0.435\textwidth}
    \includegraphics[width=1.0\linewidth]{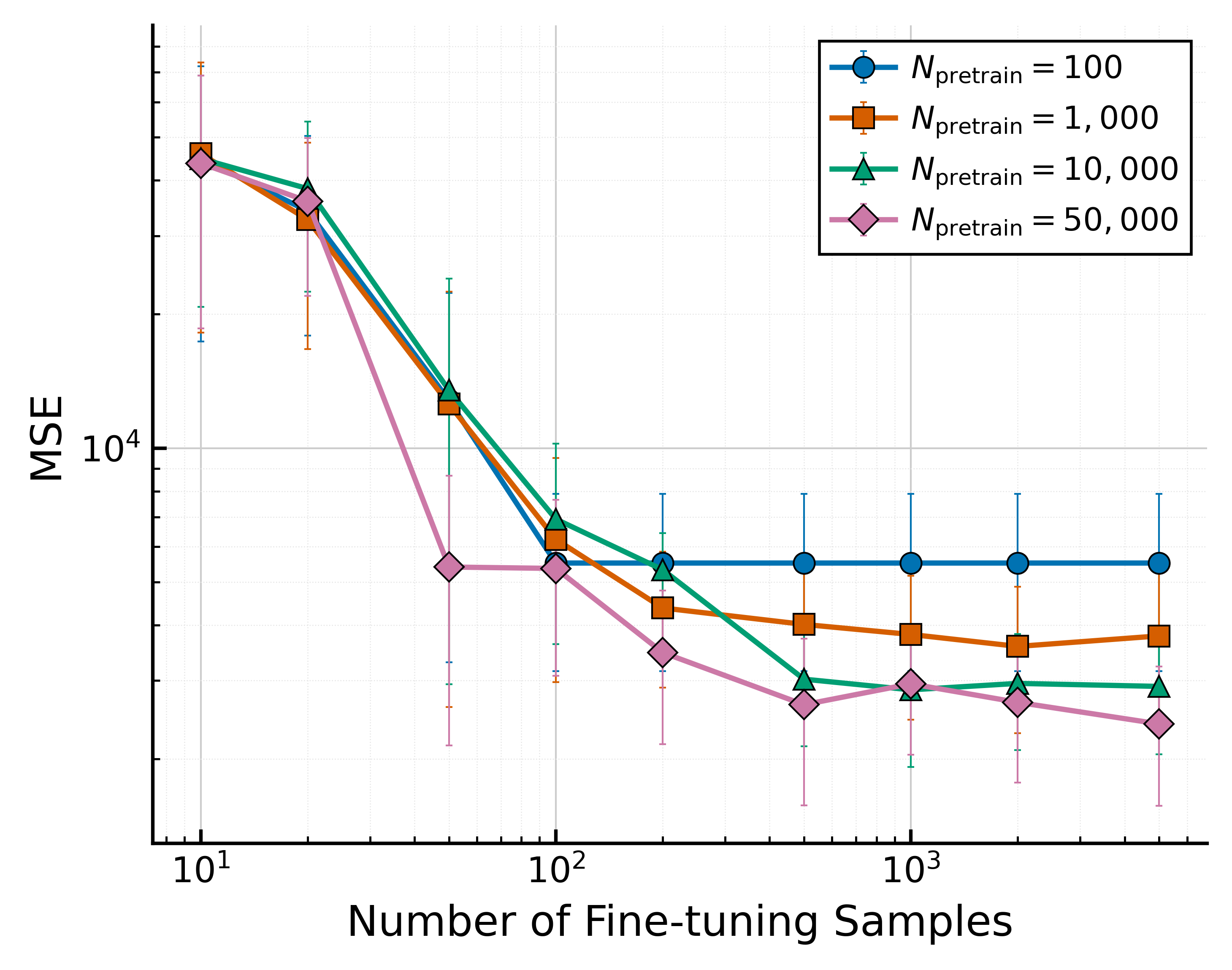}%
    \end{minipage}
    \caption{Trispecific antibody optimization and transfer learning efficiency. The left side illustrates the two topological variants evaluated, where the top configuration (format A) represents the safe format and the bottom one (format B) represents the toxic phenotype driven by distal domain placement. The center plot shows the distribution of the training data activity values for both formats. The right panel plots the model's test MSE against the number of fine-tuning samples, demonstrating that pretraining on larger datasets significantly improves predictive accuracy in data-sparse regimes.}
    \label{fig:trispecific}
\end{figure}

Figure~\ref{fig:trispecific} (left) depicts the graph representations of these two trispecific formats, where the therapeutic index is governed by molecular geometry rather than simple affinity. As defined in the ground-truth landscape, the configuration on the left corresponds to the safe therapeutic format, whereas the architecture on the right represents the toxic phenotype induced by the specific distal placement of the high-affinity domain. We leveraged this setup to evaluate a transfer learning workflow, pretraining the model on simple, single-domain affinity data (i.e. intrinsic node values) before fine-tuning on the complex, geometry-dependent task. The quantitative benefit of this approach is detailed in Figure~\ref{fig:trispecific} (right), which plots the Test Mean Squared Error (MSE) as a function of available complex-format fine-tuning samples. The performance curves revealed a clear hierarchy based on pretraining volume: models pretrained on large datasets of related monospecifics (up to $N_\text{pretrain} = 50,000$) achieve significantly lower error rates when fine-tuning data is scarce, effectively bridging the data gap inherent to multispecific engineering.

\begin{figure}
    \centering
    \includegraphics[width=0.35\linewidth]{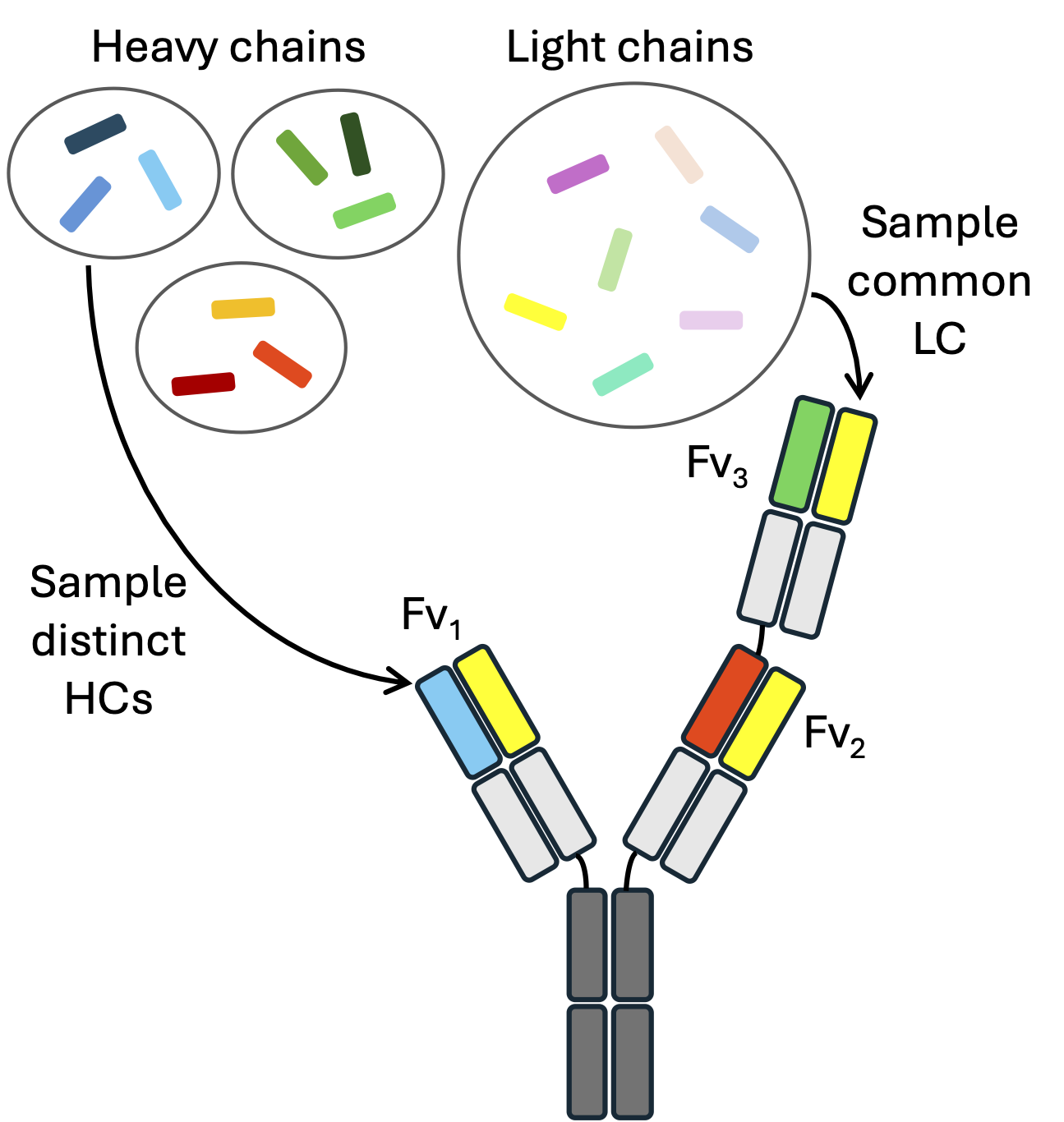}\quad%
    \includegraphics[width=0.5\linewidth]{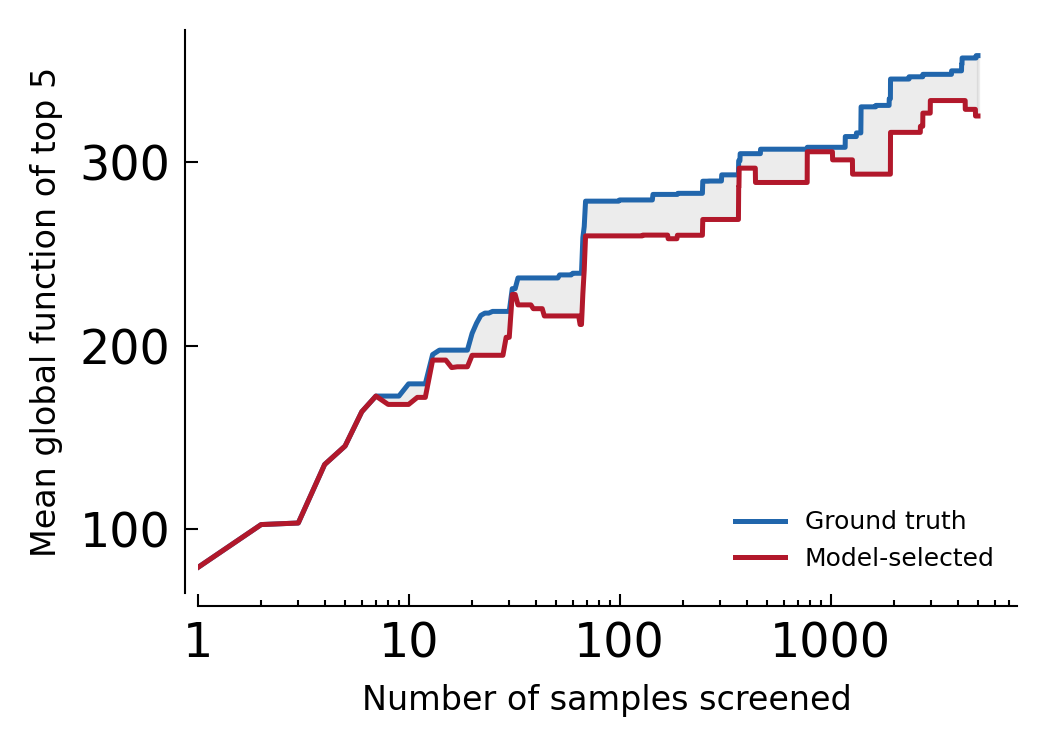}%
    \caption{Left: Identification of best common light chain from sampling combinations of domains. Right: Mean global function value of the top 5 trispecific constructs as a function of the number of light chain candidates considered (sampled from the combined 75k pretraining pool for each format). The ground truth curve represents the theoretical optimum found selecting according to actual Ehrlich function values. The model-selected curve shows the average true value of the 5 candidates ranked highest by the fine-tuned graph neural network. The close alignment indicates the model effectively ranks CLCs that maximize topology-dependent efficacy.}
    \label{fig:clc}
\end{figure}

Beyond geometric optimization, we evaluated the model's ability to identify optimal common light chains (CLCs) to improve developability and manufacturability properties. 
We utilized the pretraining data to construct distinct heavy chain libraries for each arm and a consolidated candidate pool of 75,000 light chains per format, decoupling the chains to simulate a library screening. Sampling is done across both formats A and B shown in Figure~\ref{fig:trispecific}.
We then performed the combinatorial search described in Section~\ref{sec:methods-clc}, tasking the best-performing fine-tuned GIN with selecting the light chains that, when paired commonly across all three heavy chain arms, maximized the predicted global function value. Figure~\ref{fig:clc} compares the mean global function value of the top 5 GIN-selected CLCs against ground truth top 5 average as the size of the search library increases. The model-selected sequences closely tracked the ground truth optimum, demonstrating it can effectively retrieve high-fitness, developable variants from large combinatorial spaces without exhaustive experimental validation.
The discovery of high-performing CLCs is vital for the clinical translation of multispecific therapeutics, as the use of a single light chain drastically reduces the combinatorial complexity of chain pairing. This strategy minimizes the formation of mispaired byproducts during expression, thereby improving manufacturing yield and purity, while potentially reducing the immunogenicity risks associated with non-native structural interfaces.

\section{Discussion}
\label{ref:discussion}

In this work, we introduced Synapse, a computational framework designed to benchmark machine learning architectures on the complex task of multispecific antibody property prediction. By formalizing these molecules as graphs and generating synthetic landscapes where function is strictly an emergent property of domain connectivity, we systematically evaluated the limitations of sequence-centric approaches in navigating the vast combinatorial design space of next-generation therapeutics.

Our scaling analysis demonstrates that while structure-agnostic models can approximate functional values for simple formats, they fundamentally fail to generalize to higher-order constructs where domain topology varies. Because sequence-only models treat the antibody as an unordered set of domains, they are theoretically incapable of resolving distinct geometries comprised of identical domains. In contrast, the GNN successfully minimized error across all complexity levels, confirming that encoding connectivity information is necessary to not only approximate topology-dependent functions such as steric shielding or avidity gating, but to learn generalized topological rules that can extend to novel molecular formats.

Our results address the challenge of data scarcity in multispecific engineering. We demonstrated that a fine-tuning strategy allows models to leverage more abundant, simple affinity data to improve performance on complex formats. As shown in our transfer learning experiments, pretraining on monospecific constructs significantly reduced the test error for trispecific predictions, effectively bridging the gap between high-throughput binding screens and low-throughput functional assays.

The practical implications of this topological awareness were highlighted in our simulated trispecific optimization case study. The GNN successfully distinguished between ``safe" (proximal) and ``toxic" (distal) configurations of TCEs, a distinction driven primarily by geometry rather than sequence composition. Furthermore, by extending this framework to the retrieval of common light chains, we showed that graph-based models can optimize developability constraints, such as minimizing mispaired byproducts, which is crucial for clinical translation.

Historically, validating such dependencies has been hindered by the lack of consistent, multi-format experimental data. While Synapse addresses this by providing a controlled ground truth, it remains a synthetic proxy. Future work must focus on validating these graph-based predictions against large-scale, wet-lab datasets of multispecific constructs to confirm that the simulated topological dependencies mirror \textit{in vivo} behavior. Nevertheless, by establishing a robust benchmarking environment, this work lays the foundation for data-efficient, topology-aware computational antibody design that moves beyond the additive assumptions of traditional engineering.

\section{Methods}
\label{sec:methods}

\subsection{Synapse Data Generation}
\label{sec:methods-data}
The Synapse framework generates a dataset of multispecific antibody graphs $\mathcal{G} = (\mathcal{V}, \mathcal{E})$, where each graph is associated with a scalar value $y$ defined through a global function $\Phi$.

\subsubsection{Intrinsic Node Function: Ehrlich Functions}
The intrinsic function $F_i$ for a given domain node $i$ is defined by an Ehrlich function~\citep{chen2025generalists}. This function computes a score for a domain sequence $\mathbf{s}$ based on the simultaneous satisfaction of $c$ predefined ``gapped motifs" $\{(\mathbf{m}^{(j)}, \mathbf{g}^{(j)}) \mid j=1...c\}$. The global function $F(\mathbf{s})$ is the product of the individual motif satisfaction scores $h_q$
\begin{equation}\label{eq:ehrlich_global}
    F(\mathbf{s}) = \prod_{j=1}^c h_q(\mathbf{s}, \mathbf{m}^{(j)}, \mathbf{g}^{(j)})\,.
\end{equation}
The quantized satisfaction score $h_q$ for a single motif is defined as
\begin{equation}\label{eq:ehrlich_motif}
 h_q(\mathbf{s}, \mathbf{m}^{(j)}, \mathbf{g}^{(j)}) = \max_{\ell < L} \frac{1}{q} \left\lfloor \frac{q}{k} \sum_{p = 1}^k \mathds{1}\{ s_{\ell + g_p^{(j)}} = m_p^{(j)}\} \right\rfloor\,,
\end{equation}
where $k$ is the motif length, $q$ is a quantization parameter, and $\mathds{1}\{\cdot\}$ is the indicator function. This product-based formulation creates a non-additive landscape with fitness cliffs.

We use a motif length of 3 and sample 10 motifs per binding domain across all numerical experiments.

\subsubsection{Sequence Sampling}
To ensure that the synthetic domain sequences $s_i$ retain the statistical properties of natural antibodies, we utilized the Observed Antibody Space (OAS) database~\citep{oas1,oas2} to construct a Position-Specific Scoring Matrix (PSSM). This PSSM, derived from AHo-aligned heavy and light chain sequences, defines the probability distribution $\mathcal{P}_{l}(a)$ of amino acid $a$ at each position $l$.

Data generation proceeded in two stages to mirror the exploration of a local fitness landscape. First, an initial ``seed" sequence is sampled directly from OAS. This sequence serves as the ground-truth motif for the node's Ehrlich function, ensuring the landscape contains a provably solvable optimum where $F(s_\text{seed}) = 1$.
Second, to generate the full dataset of training samples, we apply a position-aware perturbation scheme to the seed sequence. For every position $l$ in the sequence, a mutation is triggered with a defined edit probability $p_\text{edit}$, set to 0.3 in numerical experiments. 
When a mutation occurs, the new residue is not chosen uniformly; instead, it is resampled from the local PSSM distribution $\mathcal{P}_{l}$. This strategy ensures that even lower-fitness variants in the dataset respect evolutionary constraints and maintain biophysically realistic sequence compositions.

\subsubsection{Global Graph Function}
We define the global biological activity $y$ of an antibody graph $\mathcal{G} = (\mathcal{V}, \mathcal{E})$ as an emergent property derived from both the intrinsic properties of its domains and their topological arrangement.

First, we establish the intrinsic value vector $\mathbf{f} \in \mathbb{R}^{n}$, where $f_i = F_i(\mathbf{s}_i)$ for variable domains and $f_i = 0$ for constant domains and $n=|\mathcal{V}|$.
A scalar global function $y$ is obtained via a global readout function $\Phi: \mathbb{R}^{n} \times \mathcal{G} \rightarrow \mathbb{R}$ that maps the state vector and the graph topology to a single scalar
\begin{equation}\label{eq:synapse_global_fn}
    y = \Phi(\mathbf{f}, \mathcal{G})\,.
\end{equation}
The definition of $\Phi$ is flexible, allowing the framework to model diverse biological landscapes. For the scaling analysis in Section~\ref{sec:scaling}, $\Phi$ was taken to be aggregate contributions from a node's local structural neighborhood by incorporating the graph adjacency matrix $\mathbf{A}$
\begin{equation}\label{eq:scaling_phi}
y = \frac{1}{n} |\mathbf{w} \odot \mathbf{f} + \mathbf{A}(\mathbf{w} \odot \mathbf{f}) + \mathbf{A}^2(\mathbf{w} \odot \mathbf{f})|\,,
\end{equation}
where $\odot$ denotes the Hadamard product and $\mathbf{w} \in \mathbb{R}^{n}$ is a randomly initialized weight vector sampled independently for each format and connectivity. The term $\mathbf{A}\mathbf{v}$ represents the influence of immediate neighbors, while $\mathbf{A}^2\mathbf{v}$ captures second-order structural dependencies. For the trispecific case study in Section~\ref{sec:trispecific}, $\Phi$ included non-linear interaction terms and topology to model avidity and conditional toxicity, as described in equation~(\ref{eq:trispecific_function}). A more general formulation of possible global function values is given in Appendix~\ref{sec:format-aware-score}.

\subsection{Graph Isomorphism Network Architecture}
\label{sec:methods-gnn}
Our primary model is a Graph Isomorphism Network (GIN)~\citep{xu2018how}. The model first generates an initial hidden state $\mathbf{h}_i^{(0)} \in \mathbb{R}^d$ for each node $i$ by embedding its amino acid sequence $s_i$. Let $\mathbf{E} \in \mathbb{R}^{|\Sigma| \times d}$ be a learnable embedding matrix for the amino acid alphabet $\Sigma$. The node initialization is computed as the mean of the residue embeddings
\begin{equation}
    \mathbf{h}_i^{(0)} = \frac{1}{|s_i|} \sum_{a \in s_i} \mathbf{E}[a]\,.
\end{equation}
These hidden states are iteratively updated for $k=1...K$ layers via the GIN update rule
\begin{equation}
    \mathbf{h}_i^{(k)} = \text{MLP}^{(k)} \left( (1 + \epsilon^{(k)}) \cdot \mathbf{h}_i^{(k-1)} + \sum_{j \in \mathcal{N}(i)} \mathbf{h}_j^{(k-1)} \right)\,,
\end{equation}
where $\mathcal{N}(i)$ is the set of neighbors of node $i$, $\text{MLP}^{(k)}$ is a 2-layer Multi-Layer Perceptron, and $\epsilon^{(k)}$ is a learnable parameter.

After $K$ layers, a graph-level representation $\mathbf{h}_{\mathcal{G}}$ is obtained by global mean pooling of the final node states
\begin{equation}
    \label{eq:gin-pool}
    \mathbf{h}_{\mathcal{G}} = \frac{1}{n} \sum_{i \in \mathcal{V}} \mathbf{h}_i^{(K)}\,.
\end{equation}
Finally, this graph representation is passed through a regression head to predict the scalar global function 
\begin{equation}
    \label{eq:gin-regression}
    \hat{G} = \text{MLP}_{\text{reg}}(\mathbf{h}_{\mathcal{G}})\,.
\end{equation}

\subsection{MLP Baseline Models}
\label{sec:methods-mlp}
To isolate the contribution of graph-based message passing, we implemented a graph-unaware MLP baseline. This model shared an identical architecture for node feature initialization ($\mathbf{h}_i^{(0)}$), global pooling ($\mathbf{h}_G$), and the final regression head ($\hat{G}$).

The key difference is the absence of message passing. Instead of the GIN update rule, the update at each layer $k$ is an independent MLP applied to each node's representation, with no information from its neighbors
\begin{equation}
    \mathbf{h}_i^{(k)} = \text{MLP}_{\text{node}}^{(k)}(\mathbf{h}_i^{(k-1)})\,.
\end{equation}
This model is therefore exposed to the same sequence information as the GIN but is blind to the graph connectivity $\mathbf{A}$. The final layer layer $K$ is then pooled and passed through a regression head, following Equations~(\ref{eq:gin-pool}) and~(\ref{eq:gin-regression}).

Additionally, we implemented a format-conditioned MLP baseline to determine if explicit topology is required or if format identity suffices. This model retained the architecture of the graph-unaware MLP but received connectivity information as a learnable embedding from the one-hot encoded format, which is concatenated to each node embedding prior to input into the MLP.

\subsection{Training and Implementation}
\label{sec:methods-training}

All graph neural network and baseline models were implemented using the PyTorch Geometric library~\citep{fey2019fast} within the PyTorch framework. The experimental pipeline, including training loops and model evaluation, was orchestrated using PyTorch Lightning~\citep{paszke2019pytorch}.

Network parameters were optimized by minimizing the Mean Squared Error (MSE) loss using the Adam optimizer~\citep{kingma2014adam}. A comprehensive summary of the model architectures, optimization hyperparameters, and training configurations is provided in Table~\ref{tab:hyperparameters}. 
For evaluation, the synthetic datasets were randomly partitioned into training ($80\%$), validation ($10\%$), and test ($10\%$) sets. 

\begin{table}[ht]
\centering
\caption{Hyperparameters used in experiments.}
\label{tab:hyperparameters}
\begin{tabular}{llc}
\toprule
\textbf{Category} & \textbf{Hyperparameter} & \textbf{Value} \\
\midrule
\multicolumn{3}{l}{\textit{Model Architecture}} \\
& Embedding dimension & 64 \\
& Number of GNN layers & 3 \\
& Hidden layer multiplier & 2 \\
& Sequence aggregation & Mean \\
\midrule
\multicolumn{3}{l}{\textit{Optimizer}} \\
& Algorithm & Adam \\
& Learning rate & 0.001 \\
\midrule
\multicolumn{3}{l}{\textit{Training}} \\
& Epochs & 50 \\
& Batch size & 16 \\
\midrule
\multicolumn{3}{l}{\textit{Synthetic Dataset}} \\
& Number of motifs & 10 \\
& Motif length & 3 \\
& Edit probability & 0.3 \\
\bottomrule
\end{tabular}
\end{table}

\subsection{Common Light Chain Retrieval}
\label{sec:methods-clc}
To simulate CLC optimization, we organized the pretraining sequences into separate libraries for heavy and light chains. Specifically, for the trispecific topology in Section~\ref{sec:trispecific}, we established for each format three distinct datasets of heavy chains ($H_1, H_2, H_3$), each containing 25,000 sequences, and collated the corresponding light chains into a single combined pool $L_\text{pool}$ of 75,000 sequences. This is done separately for both format A and format B, which differ by a swap of the Ehrlich functions between the first and third node, and in the global score function, as given in Equation~(\ref{eq:trispecific_function}). For a construct with fixed heavy chains $s_{h_1} \in H_1, s_{h_2} \in H_2, s_{h_3} \in H_3$, we enforced a constraint where all nodes share a common light chain sequence $l \in L_\text{pool}$. The optimization objective was to identify $s_{l^*} = \text{argmax}_{s_l \in L_\text{subset}} \Phi(\mathbf{F}(\mathbf{s}_h, s_l), \mathcal{G})$, where $\mathbf{F}$ computes intrinsic Ehrlich values for the re-paired domains. We performed this retrieval over increasing subsets of $L_\text{pool}$ sampled randomly across both trispecific formats, comparing the best global function value $y$ obtained when ranking candidates via ground-truth Ehrlich functions versus the fine-tuned GIN predictions.

\section*{Data availability}
Antibody sequences used to fit Synapse were obtained from the OAS database~\cite{oas1,oas2}.
All code related to this study is available at \url{https://github.com/prescient-design/synapse}.

\section*{Acknowledgements}
We thank Mercedesz Balazs, Edith Lee, Tyler Bryson, Simon Kelow, Andrew Leaver-Fay, Margaret Porter Scott, Hubert Kettenberger, Robert Herrera, Hok Seon Kim, Wen-ting Tsai and Christoph Spiess for useful discussions.

\appendix
\section{Format-aware scoring function}
\label{sec:format-aware-score}

We define a \emph{format-aware scoring function} that maps a multi-specific antibody complex to a scalar value by explicitly coupling per-domain functional values with the complex's molecular connectivity.
Each complex is represented as an undirected graph $\mathcal{G} = (\mathcal{V}, \mathcal{E})$ with \(n=|\mathcal{V}|\) nodes corresponding to binding domains and adjacency matrix \(\mathbf{A}\in\{0,1\}^{n\times n}\) determined by the selected connectivity format.
Let \(\mathbf{f}\in\mathbb{R}^n\) denote the per-node functional values, where binding nodes are assigned closed-form functional scores and constant nodes are initialized to zero.

To enable non-zero interactions involving constant domains while maintaining numerical stability, we construct a stabilized interaction vector
\begin{equation}
\qquad
\mathbf{f}' = \mathrm{sigmoid}\!\left(
\frac{\mathbf{f} - \mu_{f}}{\sigma_{f} + \delta}
\right),
\end{equation}
where \(\delta >0\) is a small constant, and \(\mu_{f},\sigma_{f}\) are the sample mean and standard deviation of \({\mathbf{v}}\).
We then define the symmetric interaction matrix \(\mathbf{M} = \mathbf{f}'(\mathbf{f}')^\top\).

\begin{figure}
    \centering
    \includegraphics[width=1.0\linewidth]{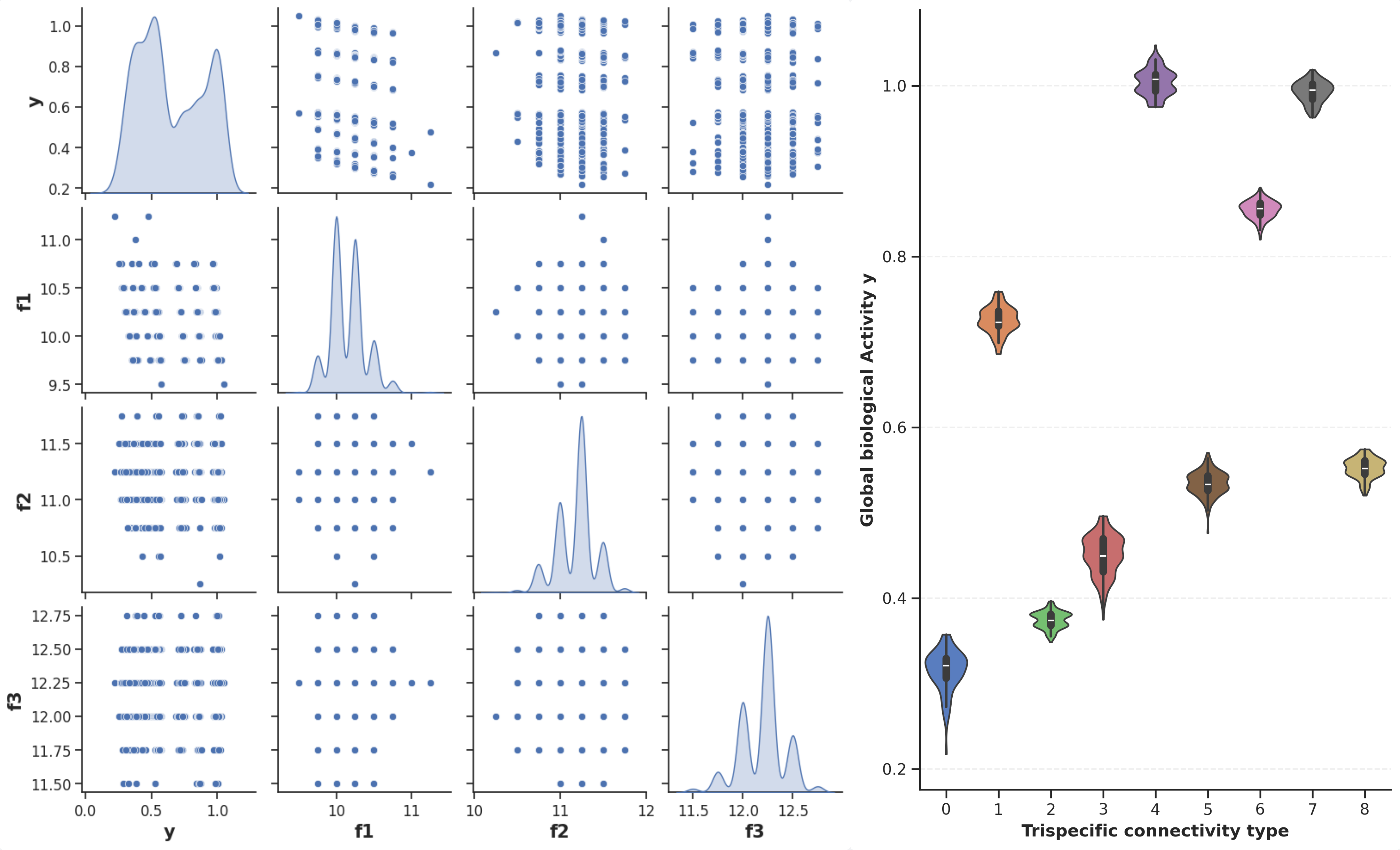}
    \caption{Characterization of trispecific antibody biological activity profiles. The univariate kernel density estimate of the global activity metric ($y$) reveals a multimodal distribution, suggesting underlying structural heterogeneity across the generated population. 
    The left panel resolves this complexity through pairwise domain correlations ($f1$-$f3$) and the right panel through a format-stratified violin plot. 
    This stratification demonstrates that distinct connectivity types (9 formats in total) occupy unique activity regimes, confirming that the scoring function successfully captures and distinguishes format-dependent distributional shifts.}
    \label{fig:format-aware}
\end{figure}

Pairwise interactions are modulated by a synergy matrix \(\mathbf{J}\in\mathbb{R}^{n\times n}\), which encodes domain-specific constraints.
In particular, all interactions involving the terminal constant node are suppressed, while interactions involving the penultimate constant node decay exponentially across its partners.
To ensure explicit format dependence, interactions are gated by the adjacency matrix via the elementwise product \(\mathbf{J}\odot \mathbf{A}\).
The resulting synergy term is
\begin{equation}
S(\mathbf{f},\mathcal{G}) = \sum_{i<j} (\mathbf{J}\odot \mathbf{A})_{ij}\, \mathbf{M}_{ij}.
\end{equation}

We further include a degree-weighted linear penalty and a weak higher-order cooperative term,
\begin{equation}
P(\mathbf{f},\mathcal{G}) = \sum_{i=1}^n \lambda_i\, d_i\, f_i, \qquad
C(\mathbf{f},\mathcal{G}) = \alpha
\prod_{i=1}^{n-2} f_i,
\end{equation}
where \(\lambda_i\ge 0\) are fixed per-node penalties, and \(\alpha\) controls the strength of the cubic interaction.
The final format-aware score is given by
\begin{equation}
y = \Phi(\mathbf{f}, \mathcal{G}) =
S(\mathbf{f},\mathcal{G}) - P(\mathbf{f},\mathcal{G})
+ \frac{C(\mathbf{f},\mathcal{G})}{\mathbb{E}[|C(\mathbf{f},\mathcal{G})|]+\delta},
\end{equation}
optionally followed by a sigmoid or softplus nonlinearity.
Because the adjacency matrix \(\mathbf{A}\) gates the interaction structure, two complexes with identical per-domain values \(\mathbf{f}\) but different molecular formats generally receive different scores. We visualize multiple connectivity trispecific formats under fixed per-domain values to isolate and demonstrate the effect of molecular arrangement on the format-aware score. We first evaluated the distribution of the objective function across all samples. Figure \ref{fig:format-aware} (left) illustrates that the global activity profile is non-Gaussian and multimodal, suggesting the presence of underlying structural subclasses. When these samples are partitioned by connectivity type (Figure \ref{fig:format-aware} right), the distributions become unimodal and distinct, confirming that the observed multimodality is driven by format-specific performance constraints rather than random noise.

\bibliography{references}

\end{document}